\documentclass[12pt]{article}
\usepackage{eurosym}
\usepackage{amssymb}
\usepackage{amsmath}
\usepackage{cite}
\usepackage{graphicx}
\usepackage{epsf}
\usepackage{lscape}
\usepackage{url}
\begin{document}

\title{Lie symmetry classification and qualitative analysis for the fourth-order Schr\"{o}dinger
equation}
\author{A. Paliathanasis$^{1,2}$\thanks{%
Email: anpaliat@phys.uoa.gr}, G. Leon$^{3,1}$ and P.G.L. Leach$^{1,4}$ \and
$~^{1}$Institute of Systems Science, Durban University of Technology \\
\ PO Box 1334, Durban 4000, Republic of South Africa\ \\
$^{2}$Instituto de Ciencias F\'{\i}sicas y Matem\'{a}ticas,\\
\ Universidad Austral de Chile, Valdivia 5090000, Chile\\
$^{3}$Departamento de Matem\'{a}ticas, Universidad Cat\'{o}lica del Norte, \\
Avda. Angamos 0610, Casilla 1280 Antofagasta, Chile \\
$^{4}$School of Mathematical Sciences, University of KwaZulu-Natal,\\
Durban, Republic of South Africa}
\maketitle

\begin{abstract}
The Lie symmetry analysis for the study of a $1+n~$fourth-order Schr\"{o}dinger equation inspired by the modification of the deformation algebra in the presence of a minimum length is applied. Specifically, we perform a detailed classification for the scalar field potential function where non-trivial Lie symmetries exist and simplify the Schr\"{o}dinger equation. Then, a qualitative analysis allows for the reduced ordinary differential equation to be analyzed to understand the asymptotic dynamics.

\bigskip

\end{abstract}

\section{Introduction}

\label{sec1}
The Lie symmetry analysis is a systematic approach to the study of
nonlinear differential equations \cite{Stephani, Bluman}. The existence of a symmetry vector for a given differential equation indicates the existence of invariant functions, then used to simplify the differential equation
and when it is possible to determine exact or analytic solutions \cite%
{Leach88,Ibrag98,ibra2,Azad,JPA2d,ref8,ref9,ref12,ref13,ref14,ref19,ref20}.
Moreover, symmetries can be used for the determination of conservation laws,
as also to identify equivalent dynamical systems \cite{nss1,nss2,nss3,nss4}.
Finally, the Lie symmetry analysis covers a wide range of applications in all areas of
applied mathematics. In this work, we are interested in the symmetry classification of a higher-order differential equation.

Consider the fourth-order partial differential equations known as Schr\"{o}dinger equation%
\begin{equation}
i\frac{\partial \Psi }{\partial t}+\alpha \Delta \Psi +\gamma \Delta
^{2}\Psi +V\left( \Psi \right) =0,  \label{sc.01}
\end{equation}%
with $\gamma \neq 0$, $\Delta $ the Laplace operator $\Delta =\frac{1}{\sqrt{%
\left\vert g\right\vert }}\frac{\partial }{\partial x^{\mu }}\left( \sqrt{%
\left\vert g\right\vert }g^{\mu \nu }\right) \frac{\partial }{\partial
x^{\nu }}$, $g_{\mu \nu }$~is the metric tensor which describes the physical
space. The fourth-order Schr\"{o}dinger equation was introduced in \cite%
{karp,karp1} in order to investigate the effects of the presence of small
fourth-order dispersion terms in the propagation of laser beams in a bulk
medium with Kerr nonlinearity. For $V\left( \Psi \right) =\left\vert \Psi
\right\vert ^{2p}\Psi $, the stability of solitons investigated by Karpman
in \cite{karp}. It was found that when $g_{\mu \nu }$ is the Euclidian
manifold, then for $p\dim \left( g\right) <4$, the soliton solutions are
stable.\ Moreover, as it has been shown in \cite{karp1}, equation (\ref%
{sc.01}) can follow by a variational principle. Since then, the fourth-order
Schr\"{o}dinger equation was the subject of study in various articles in the
literature, see for instance \cite{vp1,vp2,vp3,vp4,vp5,vp6,vp7,vp8}.

However, equation (\ref{sc.01}) describes also the modified Schr\"{o}dinger
equation for a particle in the context of the Generalised Uncertainty Principle
(GUP). GUP has its origin in the existence of a minimal length of the order
of the Planck length ($l_{PL}$). The latter is a standard prediction of
different approaches to quantum physics and gravity, that is, from string
theory, noncommutative geometry, and others \cite{ml1,ml2,ml3,ml4}. \
Specifically, the minimal length in Heisenberg's Uncertainty
Principle \cite{Maggiore} is introduced. For a review on GUP, we refer the reader in \cite%
{sb1}.

In the simplest case of quadratic GUP, the modified Heisenberg's Uncertainty
Principle reads
\begin{equation}
\Delta X_{\mu }\Delta P_{\nu }\geqslant \frac{\hbar }{2}[\delta
_{ij}(1+\beta P^{2})+2\beta P_{\mu }P_{\nu }]. \label{sc.02}
\end{equation}%
Consequently, the deformed algebra follows \cite{Vagenas,Moayedi},
\begin{equation}
\lbrack X_{\mu },P_{\nu }]=i\hbar \lbrack \delta _{ij}(1-\beta P^{2})-2\beta
P_{\mu }P_{\nu }],  \label{xp-com}
\end{equation}%
where $\beta $ is the parameter of deformation defined by $\beta ={\beta _{0}%
}/{M_{Pl}^{2}c^{2}=\beta }_{0}\ell _{Pl}^{2}/2\hbar ^{2}$, where $M_{Pl}$ is
the Planck mass, $\ell _{Pl}$ ($\approx 10^{-35}~m)$ is the Planck length
and $M_{Pl}c^{2}$ ($\approx 1.2~\times ~10^{19}~GeV)$ the Planck energy,
such that $\beta ^{2}\rightarrow 0$. Thus, we can consider the coordinate
representation of the modified momentum operator is $P_{\mu }=p_{\mu
}(1-\beta p^{2})$ \cite{Moayedi}, while keeping $X_{\mu }=x_{\mu }$
undeformed.

Thus the time-independent Schr\"{o}dinger equation reads%
\begin{equation}
\left( g^{\mu \nu }P_{\mu }P_{\nu }-\left( mc\right) ^{2}\right) \Psi =0.
\label{sc.03}
\end{equation}%
That is,%
\begin{equation}
-2\beta \hbar ^{2}\Delta ^{2}\Psi +\Delta \Psi +\left( \frac{mc}{\hbar }%
\right) ^{2}\Psi =0,\label{sc.04}
\end{equation}%
where we have assumed terms with $\beta ^{2}\rightarrow 0$. The fourth-order
equation (\ref{sc.04}) is the static version of (\ref{sc.01}) for $V\left(
\Psi \right) $ be a linear function. For some recent applications of GUP in
physical theories see \cite{gp1,gp2,gp3,gp4,gp5} and references therein.

In the following, we perform a complete classification of function $V\left(
\Psi \right) $ according to the admitted Lie point symmetries of equation (%
\ref{sc.01}). Such a classification scheme was proposed in the previous century
by Ovsiannikov, where the Lie point symmetries for the nonlinear equation $%
u_{t}=\left( f\left( u\right) u_{x}\right) _{x}$ were classified \cite{ov0}, leading to new interesting problems in applied mathematics and physics  \cite{ov1,ov2,ov3,ov4,ov5,ov6,ov7,ov8}.
Apart from the analysis of symmetries, the concept of asymptotic solution and boundary layer is essential in our context \cite{Verhulst}.

The plan of the paper follows. In Section \ref{sec2}, we present the basic properties and definitions for
the theory of Lie symmetries of differential equations, and we introduce the concept of the boundary layer. In Section \ref{sec3}
we present our classification scheme for the Lie point symmetries of the
fourth-order Schr\"{o}dinger equation. We present some applications of the
Lie point symmetries for the construction of similarity solutions in Section %
\ref{sec4}. Finally, in Section \ref{sec5}, we summarise our results.

\section{Preliminaries}

\label{sec2}

A differential equation may be considered as a function $%
H=H(x^{i},u^{A},u_{,i}^{A},u_{,ij}^{A},...)$ in the space $%
B=B(x^{i},u^{A},u_{,i}^{A},u_{,ij}^{A},...)$, where $x^{i}$ are the
independent variables and $u^{A}$ are the dependent variables. In our
consideration for equation (\ref{sc.01}) $x^{i}=\left( t,x^{\mu }\right) $
and $u^{A}\left( x^{i}\right) =\Phi \left( x^{i}\right) .$

\subsection{Lie symmetry vector}

Consider now, the infinitesimal transformation
\begin{align}
\bar{x}^{i}& =x^{i}+\varepsilon \xi ^{i}(x^{k},u^{B})~,  \label{pr.01} \\
\bar{u}^{A}& =\bar{u}^{A}+\varepsilon \eta ^{A}(x^{k},u^{B})~,  \label{pr.02}
\end{align}%
with generator the vector field%
\begin{equation}
\mathbf{X}=\xi ^{i}(x^{k},u^{B})\partial _{x^{i}}+\eta
^{A}(x^{k},u^{B})\partial _{u^{A}}~.  \label{pr.03}
\end{equation}

The generator $\mathbf{X}$ of the infinitesimal transformation (\ref{pr.01}%
)-(\ref{pr.02}) is a Lie point symmetry for the function $H$ if there exists a
function $\lambda $ such that the following condition holds \cite%
{Stephani,Bluman}
\begin{equation}
\mathbf{X}^{[N]}(H)=\lambda H~,~\mod H=0,  \label{pr.04}
\end{equation}%
where
\begin{equation}
\mathbf{X}^{[N]}=\mathbf{X}+\eta _{\left[ i\right] }^{A}\partial
_{u_{i}^{A}}+\eta _{\left[ ij\right] }^{A}\partial _{u_{ij}^{A}}+...+\eta _{%
\left[ ij...j_{N}\right] }^{A}\partial _{u_{ij...j_{N}}^{A}}  \label{pr.05}
\end{equation}%
is the $n^{th}$ prolongation vector%
\begin{equation}
\eta _{\left[ i\right] }^{A}=\eta _{,i}^{A}+u_{,i}^{B}\eta _{,B}^{A}-\xi
_{,i}^{j}u_{,j}^{A}-u_{,i}^{A}u_{,j}^{B}\xi _{,B}^{j}~,  \label{pr.06}
\end{equation}%
with
\begin{align}
\eta _{\left[ ij\right] }^{A}& =\eta _{,ij}^{A}+2\eta
_{,B(i}^{A}u_{,j)}^{B}-\xi _{,ij}^{k}u_{,k}^{A}+\eta
_{,BC}^{A}u_{,i}^{B}u_{,j}^{C}-2\xi _{,(i|B|}^{k}u_{j)}^{B}u_{,k}^{A}  \notag
\\
& -\xi _{,BC}^{k}u_{,i}^{B}u_{,j}^{C}u_{,k}^{A}+\eta
_{,B}^{A}u_{,ij}^{B}-2\xi _{,(j}^{k}u_{,i)k}^{A}-\xi _{,B}^{k}\left(
u_{,k}^{A}u_{,ij}^{B}+2u_{(,j}^{B}u_{,i)k}^{A}\right),  \label{pr.07}
\end{align}%
and in general%
\begin{equation}
\eta _{\left[ ij...j_{N}\right] }^{A}=D_{j_{n}}\left( \eta
_{ij...j_{n-1}}^{A}\right) -u_{ij...k}^{A}D_{j_{N}}\xi ^{k}.
\end{equation}

The existence of a Lie point symmetry in a given differential equation is essential in simplifying the differential equation through the similarity transformations.
Indeed from a specific Lie symmetry vector one may defines the following
Lagrange system
\begin{equation}
\frac{dx^{i}}{\xi ^{i}}=\frac{du^{A}}{\eta ^{A}}=\frac{du_{i}^{A}}{\eta _{%
\left[ i\right] }^{A}}=\frac{du_{ij}^{A}}{\eta _{\left[ ij\right] }^{A}}=...
\end{equation}%
whose solution provides the characteristic functions $W^{\left[ 0\right]
}\left( x^{k},u^{A}\right) $,~$W^{\left[ 1\right] }\left(
x^{k},u^{A},u_{i}^{A}\right) $, etc. These functions can be used to define
the corresponding similarity transformation.

\subsection{The Concept of a Boundary Layer}
Consider the function $\psi_\varepsilon(\tau)$ defined on a domain $D\subseteq \mathbb{R}^n$ (in our case $\tau \in D \subset \mathbb{R}$) and $\varepsilon$ is a small parameter. Suppose there exists a connected subset $S\subset D$ of dimension $\leq  n$, with the property that $\psi_\varepsilon(\tau)$ has no regular expansion in each subset of $D$ containing points of $S$. Then, a neighbourhood of $S$ in $D$ with a size to be determined is a boundary layer of the function $\psi_\varepsilon(\tau)$. A boundary layer, corresponding with the subset $S$, has been found near the boundary point $\tau=0$ of the domain \cite{Verhulst}. A fundamental technique to study the behaviour of a function $\psi_\varepsilon(\tau)$ in a boundary layer is to use a local analysis. Suppose that near a point $\tau_0 \in S$ the boundary layer is characterised in
size by an order function $\delta(\varepsilon)$. We ``rescale'' or ``stretch'' the variable $\tau$ by
introducing the local variable $\xi= \frac{\tau-\tau_0}{\delta(\varepsilon)}$.
If $\delta(\varepsilon)=o(1)$, we call $\xi$ a local (stretched or boundary layer) variable. The
function $\psi_\varepsilon(\tau)$ transforms to
$\psi_\varepsilon(\tau) = \psi_\varepsilon(\tau)(\tau_0 + \delta(\varepsilon)\xi)= \phi^*_\varepsilon (\xi).$
It is then natural to continue the local analysis by expanding the function $\phi^*_\varepsilon (\xi)$
with respect to the local variable $\xi$; with hope to find again a regular expansion.
To be more precise, assume that $\phi^*_\varepsilon = O_s(1)$  (that is, $O(1)$ and $\neq o(1)$ as $\varepsilon\rightarrow 0$) near $\xi=0$.
We wish to find local approximations of $\phi^*_\varepsilon$ by a regular expansion of the form
$\phi^*_\varepsilon(\xi)= \sum_n \delta_n^*(\varepsilon) \psi_n(\xi)$ with $\delta_n^*(\varepsilon)$, $n = 0,1,2,\ldots$ an asymptotic sequence.

In many problems, the function $\psi_\varepsilon(\tau)$ has been implicitly defined as the solution of a system of differential equations with initial and/or boundary conditions. For example, suppose that we have to study the perturbation problem $L_\varepsilon \phi = f(\tau), \tau \in D + \text{other conditions}$.
$L_\varepsilon$ is an operator containing a small parameter $\varepsilon$. For instance, $L_\varepsilon = \varepsilon \frac{d^2}{d  \tau^2} + \frac{d}{d  \tau}$, $D = [0,1]$, $f(\tau) = 0$ and boundary conditions $\phi(0)=1$, $\phi(1) = 0$.
The function $\tilde{\phi}(\tau)$ will be called a formal approximation or formal expansion of $\psi_\varepsilon(\tau)$ if $\tilde{\phi}$ satisfies the boundary conditions to a certain approximation, and if $L_\varepsilon (\tilde{\phi}(\tau))= f(\tau) + o(1)$. In practise, the requirement of
$\tilde{\phi}$ to satisfy the boundary conditions in full is sometimes relaxed to $\tilde{\phi}$ to satisfy the boundary conditions to a certain approximation.
To prove that if $\tilde{\phi}$ is a formal approximation, it also is an asymptotic approximation of $\phi$ is, in general, a complex problem. Moreover, one can give straightforward and realistic examples that are not true.

\section{Symmetry classification for the fourth-order Schr\"{o}dinger
equation}

\label{sec3}

Before we proceed with the symmetry classification, we set without loss of
generality $\gamma =1$ and by a change of transformation on the variable $t$
we can remove the coefficient $i$. Hence, equation (\ref{sc.01}) can be
written in the equivalent form%
\begin{equation}
\frac{\partial \Psi }{\partial t}+\alpha \Delta \Psi +\Delta ^{2}\Psi
+V\left( \Psi \right) =0.  \label{sc.05}
\end{equation}

Moreover, with the use of the new variable $\Phi =\Delta \Psi $, the
fourth-order differential equation (\ref{sc.05}) is written as the following
Schr\"{o}dinger-Poisson system%
\begin{eqnarray}
\frac{\partial \Psi }{\partial t}+\Delta \Phi +\alpha \Phi +V\left( \Psi
\right) &=&0,  \label{sc.06} \\
\Phi -\Delta \Psi &=&0.  \label{sc.07}
\end{eqnarray}

Assume now the generic vector field
\begin{equation}
X=\xi ^{t}\left( t,x^{\mu }.\Psi ,\Phi \right) \partial _{t}+\xi ^{\mu
}\left( t,x^{\mu },\Psi ,\Phi \right) \partial _{\mu }+\eta ^{\Psi }\left(
t,x^{\mu },\Psi ,\Phi \right) \partial _{\Psi }+\eta ^{\Phi }\left( t,x^{\mu
},\Psi ,\Phi \right) \partial _{\Phi },  \label{sc.08}
\end{equation}%
where in order to be the generator of an one-parameter point transformation
in the space of variables $\left\{ x^{\mu },\Psi \right\} ,$ it should be $%
\xi _{,\Phi }^{t}=0$, $\xi _{,\Phi }^{\mu }=0$ and $\eta _{,\Phi }^{\Psi }=0$.

The $2^{nd}$ prolongation vector reads%
\begin{equation}
X^{\left[ 2\right] }=X+\eta _{\left[ t\right] }^{\Psi }\partial _{\Psi
_{t}}+\eta _{\left[ \mu \right] }^{\Psi }\partial _{\Psi _{\mu }}+\eta _{%
\left[ t\right] }^{\Phi }\partial _{\Phi _{t}}+\eta _{\left[ \mu \right]
}^{\Phi }\partial _{\Phi _{\mu }}+\eta _{\left[ \mu \nu \right] }^{\Psi
}\partial _{\Psi _{\mu \nu }}+\eta _{\left[ \mu \nu \right] }^{\Phi
}\partial _{\Phi _{\mu \nu }}.
\end{equation}

Consequently, we apply the symmetry condition (\ref{pr.04}), and by using the geometric approach described in \cite{sn1}, we summarise the classification scheme in the following theorem.

\textbf{Theorem 1:} \textit{The generic Lie point symmetry vector for the
Schr\"{o}dinger-Poisson system (\ref{sc.06}), (\ref{sc.07}) in an arbitrary
background space }$g_{\mu \nu }$\textit{, and for arbitrary function }$%
V\left( \Psi \right) $\textit{\ is }%
\begin{equation}
X_{G}=a_{1}\partial _{t}+a_{\sigma }\mathbf{K}\left( x^{\kappa }\right)
\partial _{\mu },
\end{equation}%
\textit{where }$\mathbf{K}\left( x^{\mu }\right) $\textit{\ is an isometry for the
metric tensor }$g_{\mu \nu }$\textit{, that is }$\left[ \mathbf{K}\left(
x^{\kappa }\right) ,g_{\mu \nu }\left( x^{\kappa }\right) \right] =0$\textit{%
. }

However for specific functional forms of the potential $V\left( \Psi \right)
$ the classification scheme is described as follows.

\textbf{Theorem 2: }\textit{Let the metric tensor }$g_{\mu \nu }\left(
x^{\kappa }\right) $\textit{\ and }$\mathbf{K}\left( x^{\kappa }\right) $\textit{\
describe the isometries of }$g_{\mu \nu }\left( x^{\kappa }\right) $\textit{%
, and }$\mathbf{H}\left( x^{\kappa }\right) $\textit{\ is a proper Homothetic vector
of }$g_{\mu \nu }\left( x^{\kappa }\right) $\textit{, i.e. }$\left[ \mathbf{H%
}\left( x^{\kappa }\right) ,g_{\mu \nu }\left( x^{\kappa }\right) \right]
=2g_{\mu \nu }\left( x^{\kappa }\right) $\textit{. Then for special
functional forms of }$V\left( \Psi \right) $\textit{\ the generic symmetry
vector for the Schr\"{o}dinger-Poisson system (\ref{sc.06}), (\ref{sc.07})
is:}

\textit{For }$\alpha \neq 0$,

\textit{\qquad I: For }$V\left( \Psi \right) =V_{0}\Psi $\textit{, the
symmetry vector is }$X_{G}^{I}=a_{1}\partial _{t}+a_{\sigma }\mathbf{K}%
\left( x^{\kappa }\right) \partial _{\mu }+a_{2}\left( \Psi \partial _{\Psi
}+\Phi \partial _{\Phi }\right) +a_{3}\left( F\left( t,x^{\kappa }\right)
\partial _{U}+F_{,\mu \nu }\left( t,x^{\kappa }\right) \partial _{\Phi
}\right) $\textit{, where }$F\left( t,x^{\kappa }\right) $\textit{\ is a
solution of the original system. The new coefficients in the vector field
indicate the linearisation of the system.}

\textit{For }$\alpha =0$,

\textit{\qquad II: For }$V\left( \Psi \right) =0$\textit{, the generic
symmetry vector is }$X_{G}^{II}=a_{1}\partial _{t}+a_{\sigma }\mathbf{K}%
\left( x^{\kappa }\right) \partial _{\mu }+a_{2}\left( \Psi \partial _{\Psi
}+\Phi \partial _{\Phi }\right) +a_{3}\left( F\left( t,x^{\kappa }\right)
\partial _{U}+F_{,\mu \nu }\left( t,x^{\kappa }\right) \partial _{\Phi
}\right) +a_{4}\left( 4t\partial _{t}+\mathbf{H}\left( x^{\kappa }\right)
\partial _{\mu }-2\Phi \partial _{\Phi }\right) $\textit{.}

\textit{\qquad III: For }$V\left( \Psi \right) =V_{0}\Psi $\textit{, the
generic symmetry vector is }$X_{G}^{III}=a_{1}\partial _{t}+a_{\sigma }%
\mathbf{K}\left( x^{\kappa }\right) \partial _{\mu }+a_{2}\left( \Psi
\partial _{\Psi }+\Phi \partial _{\Phi }\right) +a_{3}\left( F\left(
t,x^{\kappa }\right) \partial _{U}+F_{,\mu \nu }\left( t,x^{\kappa }\right)
\partial _{\Phi }\right) +a_{4}\left( 4t\partial _{t}+\mathbf{H}\left(
x^{\kappa }\right) \partial _{\mu }-2\Phi \partial _{\Phi }-4V_{0}t\left(
\Psi \partial _{\Psi }+\Phi \partial _{\Phi }\right) \right) $\textit{.}

\textit{\qquad IV: For}$~V\left( \Psi \right) =V_{0}\Psi ^{P+1},~P\neq -1,0$%
\textit{, the generic symmetry vector is }$X_{G}^{IV}=a_{1}\partial
_{t}+a_{\sigma }\mathbf{K}\left( x^{\kappa }\right) \partial _{\mu
}+a_{4}\left( 4t\partial _{t}+\mathbf{H}\left( x^{\kappa }\right) \partial
_{\mu }-2\Phi \partial _{\Phi }-\frac{4}{P}\left( \Psi \partial _{\Psi
}+\Phi \partial _{\Phi }\right) \right) $\textit{.}

\textit{\qquad V: For }$V\left( \Psi \right) =V_{0}\exp \left( P\Psi \right)
,~P\neq 0$\textit{, the generic symmetry vector is }$X_{G}^{IV}=a_{1}%
\partial _{t}+a_{\sigma }\mathbf{K}\left( x^{\kappa }\right) \partial _{\mu
}+a_{4}\left( 4t\partial _{t}+\mathbf{H}\left( x^{\kappa }\right) \partial
_{\mu }-2\Phi \partial _{\Phi }-\frac{4}{P}\left( \partial _{\Psi }\right)
\right) $\textit{.}

We proceed with our analysis by considering specific metric tensor $g_{\mu \nu }$.

\section{Application}

\label{sec4}

Consider now that the metric tensor $g_{\mu \nu }$ is maximally symmetric and
admit a homothetic vector field. Hence, $g_{\mu \nu }$ is necessary for the flat
space. For simplicity of our calculations, assume further that $\dim g_{\mu
\nu }=1$. The one-dimensional flat space with line element $ds^{2}=dx^{2}$
admits the isometry $\partial _{x}$ and the proper Homothetic field $%
x\partial _{x}$.

Therefore the Schr\"{o}dinger-Poisson system reads%
\begin{eqnarray}
\frac{\partial \Psi }{\partial t}+\frac{\partial ^{2}\Phi }{\partial x^{2}}%
+\alpha \Phi +V\left( \Psi \right) &=&0, \\
\Phi -\frac{\partial ^{2}\Psi }{\partial x^{2}} &=&0.
\end{eqnarray}

In the case where $\alpha \neq 0$, the generic vector field is $%
X^{I}=a_{1}\partial _{t}+a_{2}\partial _{x}$, for arbitrary potential
function $V\left( \Psi \right) $. From the elements of $X^{I}$ we can reduce
the dynamical system into the static and the stationary cases. However, from
the vector field $\partial _{t}+c\partial _{x}$ we reduce the dynamical
system as follow%
\begin{eqnarray}
-c\frac{\partial \Psi }{\partial \xi }+\frac{\partial ^{2}\Phi }{\partial
\xi ^{2}}+\alpha \Phi +V\left( \Psi \right) &=&0,  \label{sc.10} \\
\Phi -\frac{\partial ^{2}\Psi }{\partial \xi ^{2}} &=&0,  \label{sc.11}
\end{eqnarray}%
where $\xi =x-ct$ is the new independent variable and $c$ describes the
speed of the travel wave. For a linear function $V\left( \Psi \right) ,$ the
closed-form solution of the system (\ref{sc.10}), (\ref{sc.11}) can be
expressed in terms of exponential functions.

However, for $V\left( \Psi \right) =V_{0}\Psi $, there exist the additional
possible reduction $\partial _{t}+c\partial _{x}+\beta \left( \Psi \partial
_{\Psi }+\Phi \partial _{\Phi }\right) $, which provides the similarity
transformation $\Psi =e^{\beta t}\psi \left( \xi \right) $, $\Phi =e^{\beta
t}\phi \left( \xi \right) $, $\xi =x-ct$ with reduced system%
\begin{eqnarray}
-c\frac{\partial \psi }{\partial \xi }+\frac{\partial ^{2}\phi }{\partial
\xi ^{2}}+\alpha \phi +\beta \psi +V_{0}\psi &=&0, \\
\phi -\frac{\partial ^{2}\psi }{\partial \xi ^{2}} &=&0.
\end{eqnarray}
Let us focus now with the case where $\alpha =0$ and assume $V\left( \Psi
\right) =V_{0}\Psi ^{P+1}$ and $V\left( \Psi \right) =V_{0}\exp \left( P\Psi
\right) $.

\subsection{Powerlaw function $V\left( \Psi\right) =V_{0}\Psi ^{P+1}, \;
P\neq 0$}
\label{SECT:4.1}

For the power-law potential function, from the vector field $\left(
4t\partial _{t}+x\partial _{x}-2\Phi \partial _{\Phi }-\frac{4}{P}\left(
\Psi \partial _{\Psi }+\Phi \partial _{\Phi }\right) \right) $ we define the
similarity transformation%
\begin{equation*}
\Psi \left( t,x\right) =\psi \left( \sigma \right) t^{-\frac{1}{P}}~,~\Phi
\left( t,x\right) =\phi \left( \sigma \right) t^{-\frac{2+P}{2P}}~,~\sigma
\left( t,x\right) =\frac{x}{t^{\frac{1}{4}}},
\end{equation*}%
and if $P\neq 0$, with reduced system
\begin{eqnarray}
\frac{\partial ^{2}\phi }{\partial \sigma ^{2}}+V_{0}\psi ^{P+1}-\frac{1}{4}%
\sigma \frac{\partial \psi }{\partial \sigma }-\frac{1}{P}\psi &=&0,
\label{eq27} \\
\phi -\frac{\partial ^{2}\psi }{\partial \sigma ^{2}} &=&0.  \label{eq28}
\end{eqnarray}
If $\phi=0$, we have
\begin{equation}
\psi= \psi_1 \sigma +\psi_0.
\end{equation}
Then, from compatibility conditions the only possible solution is the
constant solution $\psi=\psi_0$ such that
\begin{equation}
V_0 \psi_0^{P+1}-\frac{\psi_0}{P}=0 \implies \psi_{0}= (P V_0)^{-1/P}.
\end{equation}
Therefore, we assume the nontrivial case $\phi\neq0$. Then, we have the
fourth-order equation
\begin{eqnarray}
\frac{\partial ^{4}\psi }{\partial \sigma ^{4}}+V_{0}\psi ^{P+1}-\frac{1}{4}%
\sigma \frac{\partial \psi }{\partial \sigma }-\frac{1}{P}\psi &=&0.
\label{Neweq27}
\end{eqnarray}
We introduce the logarithmic independent variable
\begin{equation}
\tau = \ln (\sigma), \label{tau-32}
\end{equation}
and redefine
\begin{align}
\psi(\sigma)= \bar{\psi}(\ln (\sigma)) \label{psi-33}
.
\end{align}
That is, for any function $f(\sigma)$, define
\begin{equation}
\bar{f}(\tau)=f(e^{\tau}).
\end{equation}
Then, using the chain rule and the relation $\sigma=e^\tau$ we obtain
\begin{align}
&\frac{\partial f }{\partial \sigma } = e^{- \tau } \bar{f }^{\prime }(\tau
), \\
&\frac{\partial ^{2} f }{\partial \sigma ^{2}} = e^{- 2 \tau } \left(\bar{f }%
^{\prime \prime }(\tau )-\bar{f }^{\prime }(\tau )\right), \\
& \frac{\partial ^{3}\psi }{\partial \sigma ^{3}}= e^{-3 \tau } \left({\bar{f%
}}^{(3)}(\tau )-3 {\bar{f}}^{\prime \prime }(\tau )+2 {\bar{f}}^{\prime
}(\tau )\right), \\
& \frac{\partial ^{4}\psi }{\partial \sigma ^{4}}=e^{-4 \tau }
\left(u^{(4)}(\tau )-6 u^{(3)}(\tau )+11 u^{\prime \prime }(\tau )-6
u^{\prime }(\tau )\right).
\end{align}
Then, \eqref{Neweq27} becomes
\begin{align}
& \frac{\bar{\psi }(\tau ) \left(P V_0 \bar{\psi }(\tau )^P-1\right)}{P}%
+\left(-6 e^{-4 \tau }-\frac{1}{4}\right) \bar{\psi }^{\prime} + e^{ -4 \tau } \bar{%
\psi }^{\prime \prime} +   e^{-4 \tau } \bar{\psi }^{(3)}(\tau )+e^{-4 \tau } \bar{%
\psi }^{(4)}(\tau )=0.  \label{Neweq36}
\end{align}
Assuming that $\bar{\psi }$ is bounded with bounded derivatives as $%
\tau\rightarrow +\infty$ we obtain the asymptotic equation
\begin{align}
\frac{\bar{\psi }_{+}(\tau ) \left(P V_0 \bar{\psi }_{+}(\tau )^P-1\right)}{P%
}-\frac{1}{4} \bar{\psi }_{+}^{\prime }(\tau )=0,  \label{eq40}
\end{align}
which admits the first integral
\begin{align}
c_1 \frac{\bar{\psi }_{+}(\tau )^P}{\left(1-P V_0 \bar{\psi }_{+}(\tau
)^P\right)}= e^{- {4 \tau }} \implies \bar{\psi }_+(\tau )=\left(P V_0+c_1
e^{4 \tau }\right){}^{-1/P}.
\end{align}
Defining
\begin{equation}
z_+(\tau):=\frac{\bar{\psi }_{+}(\tau )^P}{\left(1-P V_0 \bar{\psi }%
_{+}(\tau )^P\right)},
\end{equation}
$z_+(\tau)$ is monotone decreasing as $\tau\rightarrow +\infty$ for $P>0$
and monotone increasing as $\tau\rightarrow +\infty$ for $P<0$. In other
words, the asymptotic states of $\bar{\psi }_{+}(\tau )$ are
\begin{align}
& \lim_{\tau \rightarrow + \infty} \bar{\psi }_{+}(\tau )= 0\; \text{ if }%
P>0, V_0>0, \\
& \lim_{\tau \rightarrow -\infty} \bar{\psi }_{+}(\tau )= \psi_{0}:= (P
V_0)^{-1/P}\; \text{ if }P>0, V_0>0,
\end{align}
and
\begin{align}
& \lim_{\tau \rightarrow + \infty} \bar{\psi }_{+}(\tau )= \psi_{0}:= (P
V_0)^{-1/P}\; \text{ if }P<0, V_0<0, \\
& \lim_{\tau \rightarrow -\infty} \bar{\psi }_{+}(\tau )= 0\; \text{ if }%
P>0, V_0>0.
\end{align}
The cases of interest are as $\tau \rightarrow + \infty$. That is, the
monotonic function $z_+$ unveils the asymptotic behaviour as $\tau
\rightarrow + \infty$.

Now, assuming that $\bar{\psi }$ is bounded with bounded derivatives as $%
\tau\rightarrow -\infty$, we obtain the asymptotic equation
\begin{align}
& -6 \bar{\psi }_{-}^{\prime }(\tau )+11 \bar{\psi }_{-}^{\prime \prime
}(\tau )-6 \bar{\psi }_{-}^{(3)}(\tau )+ \bar{\psi }_{-}^{(4)}(\tau )=0,
\label{eq36}
\end{align}
with solution
\begin{align}
\bar{\psi }_{-}(\tau )= c_2 e^{\tau }+\frac{1}{2} c_3 e^{2 \tau }+\frac{1}{3}
c_4 e^{3 \tau }+c_5, \label{solution_48}
\end{align}
such that
\begin{equation}
\lim_{\tau\rightarrow -\infty } \bar{\psi }_{-}(\tau )= c_5.
\end{equation}
Substituting $\psi(\tau)= \bar{\psi }_{-}(\tau )$ in \eqref{Neweq36}, and
taking limit $\tau\rightarrow -\infty$, we obtain the compatibility
condition
\begin{equation}
c_5 \left(-1+P V_0 c_5{}^P\right)=0.
\end{equation}
That is, $c_5 \in\left\{ 0, (P  V_0)^{-1/P}\right\}.$ The choice $c_5= (P
V_0)^{-1/P}$ gives the proper matching condition
\begin{equation}
\lim_{\tau\rightarrow -\infty} \bar{\psi }_+(\tau )= \lim_{\tau\rightarrow
-\infty} \bar{\psi }_-(\tau )=(P V_0)^{-1/P}.
\end{equation}

Summarising, as $\tau\rightarrow -\infty$, $\psi(\tau) \approx \bar{\psi }%
_{-}(\tau )$, whereas, as $\tau\rightarrow +\infty$, $\psi(\tau) \approx
\bar{\psi }_{+}(\tau )$.

Let be define the new time variable $s=(1+ \tanh(\tau))/2$ that brings the
interval $(-\infty, \infty)$ to $(0, 1)$. Then, the original layer problem
becomes a two-point problem, with endpoints $0$ and $1$. The asymptotic
solutions can be found as
\begin{equation}
\Phi_{-}(s )= \bar{\psi}_-(-\text{arctanh}(1-2 s)),
\end{equation}
that is,
\begin{align}
\Phi_{-}(s )=(P V_0)^{-1/P}+\frac{c_3 s}{2-2 s}+\left(c_2 \left(\frac{1}{s}%
-1\right)+\frac{c_4}{3}\right) e^{-3 \text{arctanh}(1-2 s)}.
\end{align}
As $s\rightarrow 0^+$, we have the asymptotic behaviour $\Phi_{-} \rightarrow (P
V_0)^{-1/P}$.

Moreover,
\begin{equation}
\Phi_{+}(s )= \bar{\psi}_+(-\text{arctanh}(1-2 s)),
\end{equation}
becomes
\begin{align}
\Phi_{+}(s )= \left(P V_0+\frac{c_1 s^2}{(1-s)^2}\right)^{-1/P},
\end{align}
such that
\begin{align}
& \lim_{s \rightarrow 1^{-}} \Phi_{+}(s)= 0\; \text{ if }P>0, V_0>0.
\end{align}
And we have the matching condition
\begin{equation}
\lim_{s \rightarrow 0^{+}} \Phi_{-}(s)= \lim_{s \rightarrow 0^{+}}
\Phi_{+}(s)=(P V_0)^{-1/P}.
\end{equation}
The next step is to introduce the stretched variables $\kappa= s/ \varepsilon
$ and $\lambda= (1-s)/\varepsilon$, and write a solution
\begin{equation}
\Phi(s, \varepsilon)= \zeta (\kappa, \varepsilon) + \eta(\lambda , \varepsilon)
\end{equation}
where
\begin{equation}
\zeta \rightarrow (P V_0)^{-1/P}\; \text{as}\; \kappa=s/\varepsilon
\rightarrow \infty
\end{equation}
and
\begin{equation}
\eta \rightarrow 0\; \text{as}\; \lambda=(1-s)/\varepsilon \rightarrow
\infty.
\end{equation}

Near $s=0$, $\eta$ and its derivatives will be asymptotically negligible, so
$d^j \Phi(s, \varepsilon)/ds^j \sim (1/\varepsilon^j)\left[d^j
\zeta(\kappa,\varepsilon)/d\kappa^j\right]$. Take, for example,
\begin{equation}
\zeta_0 (\kappa, \varepsilon)= (P V_0)^{-1/P}+\frac{c_3 \kappa \varepsilon }{%
2-2 \kappa \varepsilon }+\left(c_2 \left(\frac{1}{\kappa \varepsilon }%
-1\right)+\frac{c_4}{3}\right) e^{-3 \text{arctanh}(1-2 \kappa \varepsilon
)}.  \label{eq61}
\end{equation}
Using the notation
\begin{align}
\bar{\psi}(\tau,\varepsilon)= \Phi (\kappa,\varepsilon), \; \kappa= \frac{%
\tanh (\tau )+1}{2 \varepsilon } \label{replacement1}
\end{align}
the approximated equation \eqref{eq36}, becomes
\begin{align}
& 6 \varepsilon (4 \kappa \varepsilon (4 \kappa \varepsilon -3)+1) \Phi
^{\prime }(\kappa,\varepsilon)  \notag \\
& +(\kappa \varepsilon -1) \Big[3 (24 \kappa \varepsilon (2 \kappa
\varepsilon -1)+1) \Phi ^{\prime \prime }(\kappa,\varepsilon)  \notag \\
& +4 \kappa (\kappa \varepsilon -1) \left(\kappa \Phi
^{(4)}(\kappa,\varepsilon)(\kappa \varepsilon -1)+3 \Phi
^{(3)}(\kappa,\varepsilon) (4 \kappa \varepsilon -1)\right)\Big]=0,
\end{align}
where primes means derivatives with respect to $\kappa$,
that admits the exact solution \eqref{eq61}.  Since we are taking $%
\varepsilon$ as an small parameter, we see that the initial layer problem is
of type
\begin{align}
& \left(-3 \Phi ^{\prime \prime }(\kappa )-4 \kappa \left(\kappa \Phi
^{(4)}(\kappa )+3 \Phi ^{(3)}(\kappa )\right)\right)  \notag \\
& +\varepsilon \left(6 \Phi ^{\prime }(\kappa )+3 \kappa \left(25 \Phi
^{\prime \prime }(\kappa )+4 \kappa \left(\kappa \Phi ^{(4)}(\kappa )+6 \Phi
^{(3)}(\kappa )\right)\right)\right)+O\left(\varepsilon ^2\right)=0.
\end{align}
Taking the expansion
\begin{align}
\Phi(\kappa )= \Phi_0(\kappa ) + \varepsilon \Phi_1(\kappa ) + \ldots \label{Expansion_1}
\end{align}
we obtain at first order
\begin{align}
-3 \Phi_0 ^{\prime \prime }(\kappa )-4 \kappa \left(\kappa \Phi_0
^{(4)}(\kappa )+3 \Phi_0 ^{(3)}(\kappa )\right)=0.
\end{align}
Hence,
\begin{equation}
\Phi_0 (\kappa )= \frac{4}{3} \sqrt{\kappa } (d_2 \kappa -3 d_1)+d_4 \kappa
+d_3.  \label{Expansion_0}
\end{equation}
At second order we have
\begin{equation}
60 d_2 \sqrt{\kappa }+6 d_4-4 \kappa ^2 \Phi_1^{(4)}(\kappa )-12 \kappa
\Phi_1^{(3)}(\kappa )-3 \Phi_1^{\prime \prime }(\kappa)=0.
\end{equation}
Hence,
\begin{align}
\Phi_1(\kappa)= 2 d_2 \kappa ^{5/2}+\frac{4}{3} d_6 \kappa ^{3/2}+d_4 \kappa
^2+d_8 \kappa -4 d_5 \sqrt{\kappa }+d_7, \label{Expansion_2}
\end{align}
and so on.
Finally we replace the leading order and second order terms \eqref{Expansion_0} and
\eqref{Expansion_2}, respectively, in \eqref{Expansion_1} with the replacement \eqref{replacement1}.

Near $s=1$, $\zeta$ and its derivatives will be asymptotically negligible, so
$d^j \Phi(s, \varepsilon)/ds^j \sim (1/\varepsilon^j)\left[d^j
\eta(\lambda,\varepsilon)/d\lambda^j\right]$. Take, for example,
\begin{equation}
\eta_0 (\lambda, \varepsilon)=\left(P V_0+c_1 e^{4 \text{arctanh}(1-2
\lambda \varepsilon )}\right)^{-1/P}.  \label{eq65}
\end{equation}
Using the notation
\begin{align}
\bar{\psi}(\tau,\varepsilon)= \Phi \left(\lambda\right), \; \lambda= \frac{%
1-\tanh (\tau )}{2 \varepsilon },
\end{align}
the approximated equation \eqref{eq40}, becomes
\begin{align}
2 V_0 \Phi (\lambda )^{P+1}+\lambda (1-\lambda \varepsilon ) \Phi ^{\prime
}(\lambda )=\frac{2 \Phi (\lambda )}{P},
\end{align}
which admits the solution \eqref{eq65}.

\subsection{Exponential function $V\left( \Psi \right) =V_{0}\exp \left(
P\Psi \right), \; P\neq 0$}

On the other hand, for the exponential potential $V\left( \Psi \right)
=V_{0}\exp \left( P\Psi \right), \; P\neq 0$, the similarity transformation
which corresponds to the vector field $\left( 4t\partial _{t}+x\partial
_{x}-2\Phi \partial _{\Phi }-\frac{4}{P}\left( \partial _{\Psi }\right)
\right) $ is
\begin{equation*}
\Psi \left( t,x\right) =\frac{\ln t}{P}+\psi \left( \sigma \right) ~,~\Phi
=t^{-\frac{1}{2}}\phi \left( \sigma \right) ~,~\sigma \left( t,x\right) =%
\frac{x}{t^{\frac{1}{4}}},
\end{equation*}%
where the reduced system is
\begin{eqnarray}
\frac{\partial ^{2}\phi }{\partial \sigma ^{2}}+V_{0}e^{P\psi }-\frac{1}{P}-%
\frac{1}{4}\sigma \frac{\partial \psi }{\partial \sigma } &=&0, \\
\phi -\frac{\partial ^{2}\psi }{\partial \sigma ^{2}} &=&0.
\end{eqnarray}
We introduce the logarithmic independent variable \eqref{tau-32} and defining  $\psi(\sigma)$ by \eqref{psi-33}.
Then, using the chain rule and the relation $\sigma=e^\tau$ we obtain
\begin{equation}
V_0 e^{P \bar{\psi }(\tau )}+\left(-6 e^{-4 \tau }-\frac{1}{4}\right) \bar{%
\psi }^{\prime} + e^{-4 \tau } \bar{\psi }^{\prime \prime} + e^{ -4 \tau } \bar{\psi }%
^{(3)}(\tau )+e^{-4 \tau } \bar{\psi }^{(4)}(\tau )-\frac{1}{P}=0.
\label{Neweq37}
\end{equation}
Assuming that $\bar{\psi }$ is bounded with bounded derivatives as $%
\tau\rightarrow +\infty$ we obtain the asymptotic equation
\begin{align}
V_0 e^{P \bar{\psi }_+(\tau )} -\frac{1}{P}-\frac{1}{4} \bar{\psi }%
_{+}^{\prime }(\tau )=0,
\end{align}
with solution
\begin{equation}
\bar{\psi }_+(\tau )= \ln \left(\left(P V_0+e^{4 \tau +c_1
P}\right){}^{-1/P}\right).
\end{equation}
Now, assuming that $\bar{\psi }$ is bounded with bounded derivatives as $%
\tau\rightarrow -\infty$, we obtain, as in section  \ref{SECT:4.1}, the asymptotic equation \eqref{eq36},
with solution \eqref{solution_48}.
Substituting $\psi(\tau)= \bar{\psi }_{-}(\tau )$ in \eqref{Neweq37}, and
taking limit $\tau\rightarrow -\infty$, we obtain
\begin{equation}
-\frac{1}{P}+V_0 e^{c_5 P}=0 \implies c_5=\ln\left[(P V_0)^{-1/P}\right].
\end{equation}
That is, we have the matching condition
\begin{equation}
\lim_{\tau\rightarrow -\infty} \bar{\psi }_+(\tau )= \lim_{\tau\rightarrow
-\infty} \bar{\psi }_-(\tau )=\ln\left[(P V_0)^{-1/P}\right].
\end{equation}
As in section \ref{SECT:4.1}, we have as $\tau\rightarrow -\infty$, $\psi(\tau) \approx \bar{\psi }%
_{-}(\tau )$, whereas, as $\tau\rightarrow +\infty$, $\psi(\tau) \approx
\bar{\psi }_{+}(\tau )$. Using the same method, we define the new time variable $s=(1+ \tanh(\tau))/2$ that brings the
interval $(-\infty, \infty)$ to $(0, 1)$. Then, the original layer problem
becomes a two-point problem, with endpoints $0$ and $1$. The asymptotic
solutions can be found as
\begin{align}
\Phi_{-}(s )=\ln\left[(P V_0)^{-1/P}\right]+\frac{c_3 s}{2-2 s}+\left(c_2 \left(\frac{1}{s}%
-1\right)+\frac{c_4}{3}\right) e^{-3 \text{arctanh}(1-2 s)}.
\end{align}
As $s\rightarrow 0^+$, we have the asymptotic behaviour $e^{\Phi_{-}} \rightarrow (P
V_0)^{-1/P}$.

Similarly, we have
\begin{align}
e^{\Phi_{+}(s )}= \left(P V_0+\frac{c_1 s^2}{(1-s)^2}\right)^{-1/P},
\end{align}
such that
\begin{align}
& \lim_{s \rightarrow 1^{-}} e^{\Phi_{+}(s)}= 0\; \text{ if }P>0, V_0>0.
\end{align}
Finally, by introducing the stretched variables $\kappa= s/ \varepsilon
$ and $\lambda= (1-s)/\varepsilon$, and write a solution
\begin{equation}
\Phi(s, \varepsilon)= \zeta (\kappa, \varepsilon) + \eta(\lambda , \varepsilon),
\end{equation}
where
\begin{equation}
\zeta \rightarrow \ln\left[(P V_0)^{-1/P}\right]\; \text{as}\; \kappa=s/\varepsilon
\rightarrow \infty,
\end{equation}
and
\begin{equation}
\eta \rightarrow 0\; \text{as}\; \lambda=(1-s)/\varepsilon \rightarrow
\infty.
\end{equation}
Then, the layer problem becomes a two-point problem, with endpoints $0$ and $1$, and we obtain the asymptotic solutions following similar approaches as in section \ref{SECT:4.1}.

\section{Conclusions}

\label{sec5}

Lie symmetry analysis is a powerful method for analysing nonlinear differential equations. In this study, was applied the Lie symmetry analysis to solve the group classification problem for a $1+n$-dimensional nonlinear higher-order Schr\"{o}dinger equation inspired by GUP.

The partial differential equation of our analysis admits an arbitrary potential function which was a constraint according to the admitted Lie point symmetries. For an arbitrary potential function, we found that the admitted Lie symmetries are the Killing vectors of the $n$-dimensional space
additionally to the vector field $\partial _{t}$. However, a new symmetry vector presented in Theorems 1 and 2 can exist for specific function forms of the potential function.

To demonstrate the application of the Lie symmetry vectors, we use the corresponding Lie invariants to define similarity transformations and reduce the partial-differential equation into an ordinary differential equation. Because of the non-linearity of the reduced equation, we studied the asymptotic dynamics and evolution.

In relation to asymptotic analysis we have obtained asymptotic solutions
\begin{align*}
\bar{\psi }_{-}(\tau ) & = c_2 e^{\tau }+\frac{1}{2} c_3 e^{2 \tau }+\frac{1}{3}
c_4 e^{3 \tau }+\left\{\begin{array}{cc}
(P V_0)^{-1/P} & \text{powerlaw function} \\
\ln\left[(P V_0)^{-1/P}\right] & \text{exponential function}
\end{array} \right., \\
\bar{\psi }_+(\tau ) & = \left\{\begin{array}{cc}
\left(P V_0+c_1
e^{4 \tau }\right){}^{-1/P} & \text{powerlaw function} \\
 \ln \left(\left(P V_0+e^{4 \tau +c_1
P}\right){}^{-1/P}\right) & \text{exponential function}
\end{array} \right.,
\end{align*}
with the proper matching condition
\begin{equation*}
\lim_{\tau\rightarrow -\infty} \bar{\psi }_+(\tau )= \lim_{\tau\rightarrow
-\infty} \bar{\psi }_-(\tau )=\left\{\begin{array}{cc}
(P V_0)^{-1/P} & \text{powerlaw function} \\
\ln\left[(P V_0)^{-1/P}\right] & \text{exponential function}
\end{array} \right..
\end{equation*}
Hence, as $\tau\rightarrow -\infty$, $\psi(\tau) \approx \bar{\psi }%
_{-}(\tau )$, whereas, as $\tau\rightarrow +\infty$, $\psi(\tau) \approx
\bar{\psi }_{+}(\tau )$.
Finally, the layer problem becomes a two-point problem, with endpoints $0$ and $1$  by introducing the stretched variables $\kappa= s/ \varepsilon
$ and $\lambda= (1-s)/\varepsilon$, and writing a formal solution
\begin{equation}
\Phi(s, \varepsilon)= \zeta (\kappa, \varepsilon) + \eta(\lambda , \varepsilon),
\end{equation}
where
\begin{equation}
\zeta \rightarrow \left\{\begin{array}{cc}
(P V_0)^{-1/P} & \text{powerlaw function} \\
\ln\left[(P V_0)^{-1/P}\right] & \text{exponential function}
\end{array} \right.,\; \text{as}\; \kappa=s/\varepsilon
\rightarrow \infty,
\end{equation}
and
\begin{equation}
\eta \rightarrow 0\; \text{as}\; \lambda=(1-s)/\varepsilon \rightarrow
\infty.
\end{equation}  Then, it is interesting to analyse possible asymptotic solutions for different initial/boundary conditions, but this numerical treatment is out of the scope of the present research. In general, when solving the problem of approximating a function $\psi_\varepsilon(\tau)$ depending on a small parameter $\varepsilon$ in a domain $D$, the following program is implemented \cite{Verhulst}.
\begin{enumerate}
    \item Try to construct a regular expansion in the original variable $\tau$. That is possible outside the boundary layers, and is usually called the outer expansion.
\item Construct in the boundary layer(s) a local expansion in an appropriate local variable. A regular expansion is usually called the inner expansion or boundary layer expansion.
\item The inner and outer expansions should be matched to obtain a formal expansion for the whole domain $D$.

\noindent
In literature, several techniques combining the three stages were developed, which makes the process more efficient. This formal expansion, which is valid in the whole domain, is sometimes called a uniform expansion. Note, however, that in the literature, expressions called  ``uniformly valid expansion" are more often than not formal expansions. So we come to the next point.
\item Prove that formal expansions, obtained in the stages 1–3, represent valid
asymptotic approximations of the function $\psi_\varepsilon(\tau)$ that we set out to study.
\end{enumerate}
This work contributes to the subject of the application of Lie point symmetries on nonlinear differential equations. In this study, we considered an Schr\"{o}dinger equation constructed by the deformation algebra of the quadratic GUP. However, that is not the unique proposed GUP, and other deformations algebras exist. Therefore, in future work, we plan to perform a detailed classification for the higher-order Schr\"{o}dinger equation for different models of GUP. Finally, we will present formal expansions, representing valid asymptotic approximations of the function $\psi_\varepsilon(\tau)$ for other initial conditions that we set out to study by singular perturbations methods, boundary layers, and multiple time scales.

\section*{Acknowledgements}
G.L. was funded by  Vicerrectoría de Investigación y Desarrollo Tecnológico at UCN.


\begin{thebibliography}{99}
\bibitem{Stephani} H. Stephani, Differential Equations: Their Solutions
Using Symmetry, Cambridge University Press, New York, (1989)

\bibitem{Bluman} G.W. Bluman and S. Kumei, Symmetries of Differential
Equations, Springer-Verlag, New York, (1989)

\bibitem{Leach88} P.G.L. Leach and V.M. Gorringe, Phys. Lett. A, 133, 289
(1988)

\bibitem{Ibrag98} R. Gazinov and N.H. Ibragimov, Nonlinear Dynamics, 17, 387
(1998)

\bibitem{ibra2} N. H. Ibragimov, On the group classification of second order
differential equations. (Russian) Dokl. Akad. Nauk SSSR, 183, 274, (1968)

\bibitem{Azad} H. Azad and M.T. Mustafa, J. Math. Anal. Appl., 333, 1180,
(2007)

\bibitem{JPA2d} M. Tsamparlis and A. Paliathanasis, \ J. Phys. A: Math.
Theor. 44, 175202 (2011)

\bibitem{ref8} F.M. Mahomed. Math. Methods Appl. Sci. 30, 1995 (2007)

\bibitem{ref9} S. Jamal, A.H. Kara,\ A.H. Bokhari, Canadian J. Phys. 90, 667
(2012)

\bibitem{ref12} A.K. Halder, A. Paliathanasis, S. Rangasamy and P.G.L.
Leach, Zeitschrift f\"{u}r Naturforschung A 74, 597 (2019)

\bibitem{ref13} S. Jamal and A.H. Kara, Nonlinear Dynamics 67, 97 (2012)

\bibitem{ref14} A.A. Chesnokov, J. Appl. Mech. Techn. Phys. 49, 737 (2008)

\bibitem{ref19} S. Jamal, Quaestiones Mathematicae 41, 409 (2018)

\bibitem{ref20} S. Jamal and N. Mnguni, Appl. Math. Comp. 335, 65 (2018)

\bibitem{nss1} A.K. Halder, A. Paliathanasis and P.G.L Leach, Symmetry 10,
744 (2018)

\bibitem{nss2} F. Schwarz, Computing 62, 1 (1999)

\bibitem{nss3} G.J. Reid and A.D. Wittkopf, ISSAC '00: Proceedings of the
2000 international symposium on Symbolic and algebraic computation, 272
(2000)

\bibitem{nss4} S. Ali, M.\ Safdar and A. Qadir, J. Applied Mathematics 2014,
793247 (2014)

\bibitem{karp} V.I. Karpman, Phys. Lett. A 215, 254 (1996)

\bibitem{karp1} V.I. Karpman and A.G. Shagalov, Physica D 144, 194 (2000)

\bibitem{vp1} J. Segata, Math. Methods Appl. Sci. 26, 1785 (2006)

\bibitem{vp2} B. Pausader, J. Functional Analysis 256, 2473 (2009)

\bibitem{vp3} B. Pausader and S. Shao, J. Hyperbolic Differential Equations
7, 651 (2010)

\bibitem{vp4} C. Baquet and E.J. Villamizar-Roa, Evolution Equations \&
Contrl Theory 9, 865 (2020)

\bibitem{vp5} X. Liu and T. Zhang, J. Math. Phys. 62, 071501 (2021)

\bibitem{vp6} B. Erdogan, W.R. Green and E.\ Torpak, J. Differential
Equations 271, 152 (2021)

\bibitem{vp7} G. Fibich, B. Ilan and G. Papanicolaou, SIAM J. Appl. Math.
62, 1437 (2002)

\bibitem{vp8} G. Fibich, B.\ Ilan and S. Schochet, Nonlinearity 16, 1809
(2003)

\bibitem{ml1} K. Konishi, G. Paffuti and P. Provero, Phys.\ Lett. B 234, 276
(1990)

\bibitem{ml2} A. Camelia, Int. J. Mod. Phys. D \textbf{11}, 35 (2002).

\bibitem{ml3} P. Martinetti, F. Mercati and L. Tomassini, Rev. Math. Phys.
24, 1250010 (2012)

\bibitem{ml4} Ashtekar, A. and Lewandowski, J., Class. Quantum Grav., 21,
R53, (2004)

\bibitem{Maggiore} M. Maggiore, Phys. Lett. B 304, 65 (1993)

\bibitem{sb1} S. Hossenfelder, Living Reviews in Relativity 16, 2, (2013)

\bibitem{Vagenas} S. Das and E.C. Vagenas, Phys. Rev. Lett. 101, 221301
(2008)

\bibitem{Moayedi} S.K. Moayedi, M.R. Setare and H Moayeri, Int. J. Theor.
Phys. 49, 2080 (2010)

\bibitem{gp1} B. Hamil, M. Merad and T. Birkandan, Eur Phys. J. Plus 134,
278 (2019)

\bibitem{gp2} M.P. Dabrowski and F.\ Wagner, EPJC 80, 676 (2020)

\bibitem{gp3} V. Nenmeli, S. Shankaranarayanan, V. Todorinov and S. Das,
Phys. Lett. B 821, 136621 (2021)

\bibitem{gp4} A. Das, S. Das and E.C. Vagenas, Phys. Lett. B 809, 135772
(2020)

\bibitem{gp5} S. Aghababaei, H. Mordpour and E.C. Vagenas, Eur. Phys. J.
Plus 136, 997 (2021)

\bibitem{ov0} L. V. Ovsiannikov, Group analysis of differential equations,
Academic Press, New York, (1982)

\bibitem{ov1} Z.-Y. Zhang and G.-F. Li, Physica A 540, 123134 (2020)

\bibitem{ov2} S. Jamal, A. H. Kara and R. Narain, J. App. Math. 2012, 765361
(2012)

\bibitem{ov3} V. Lahno, R. Zhdanov and O. Magda, Acta Appl. Math. 91, 253
(2006)

\bibitem{ov4} V.A. Baikov, A.V. Gladkov and R.J. Wiltshire, J. Phys. A:
Math. Gen. 31, 7483 (1998)

\bibitem{ov5} D. Huang and N.M. Ivanova, J. Math. Phys., 48, 073507 (2007)

\bibitem{ov6} R. Cherniha, M.\ Serov and Y. Prystavka, Comm. Nonl. Sci. Num.
Sim. 92, 105466 (2021)

\bibitem{ov7} S. Jamal, A. H. Kara and A. H. Bokhari, Can. J. Phys. 90, 667
(2012)

\bibitem{ov8} M. Tsamparlis and A. Paliathanasis, J. Phys.A: Math. Theor.
44, 175202 (2011)

\bibitem{Verhulst} Ferdinand Verhulst,
Methods and Applications of Singular Perturbations: Boundary Layers and Multiple Timescale Dynamics, Texts in Applied Mathematics, \url{https://doi.org/10.1007/0-387-28313-7}. Springer-Verlag, New York (2005)

\bibitem{sn1} A. Paliathanasis and M. Tsamparlis, J. Geom. Phys. 107, 45
(2016)

\end{thebibliography}
\end{document}